\documentclass[11pt, a4paper, logo, copyright, nonumbering]{klear}
\usepackage[authoryear, compress, round]{natbib}
\usepackage{dblfloatfix}
\usepackage{caption}
\usepackage{dramatist}
\usepackage{xspace}
\usepackage{pifont} 
\usepackage{multirow}
\usepackage{tcolorbox}
\usepackage{xltabular}
\usepackage{longtable}
\usepackage{hyperref}
\usepackage{amsfonts}
\usepackage{amsmath}
\usepackage{amssymb}
\usepackage{lineno}
\usepackage{multirow}
\usepackage{adjustbox}

\usepackage[bottom]{footmisc}

\usepackage{CJKutf8}
\usepackage{subfigure}
\usepackage{setspace}

\usepackage{dsfont}
\usepackage{array} %
\usepackage{tabularx} %
\usepackage{subfigure} %
\usepackage{xcolor} %
\usepackage{algorithm}
\usepackage{algorithmic}

\usepackage{lipsum}  %
\usepackage{multicol} %

\usepackage{graphicx}
\usepackage{booktabs}
\usepackage{multirow}

\definecolor{keywordcolor}{rgb}{0.7, 0.1, 0.1}   %
\definecolor{tacticcolor}{rgb}{0.0, 0.1, 0.6}    %
\definecolor{commentcolor}{rgb}{0.4, 0.4, 0.4}   %
\definecolor{symbolcolor}{rgb}{0.0, 0.1, 0.6}    %
\definecolor{sortcolor}{rgb}{0.1, 0.5, 0.1}      %
\definecolor{attributecolor}{rgb}{0.7, 0.1, 0.1} %

\usepackage{listings}

\usepackage{textcomp}

\usepackage{tcolorbox}
\newtcolorbox{promptbox}[1][]{
    colback=gray!10,      %
    colframe=black!50,     %
    boxrule=0.5mm,        %
    arc=1mm,              %
    boxsep=0mm,
    fontupper=\ttfamily\scriptsize,  %
    width=\textwidth,     %
    title=#1,
fonttitle=\ttfamily\footnotesize\centering,
}

\makeatletter
\def\@BTrule[#1]{%
  \ifx\longtable\undefined
    \let\@BTswitch\@BTnormal
  \else\ifx\hline\LT@hline
    \nobreak
    \let\@BTswitch\@BLTrule
  \else
     \let\@BTswitch\@BTnormal
  \fi\fi
  \global\@thisrulewidth=#1\relax
  \ifnum\@thisruleclass=\tw@\vskip\@aboverulesep\else
  \ifnum\@lastruleclass=\z@\vskip\@aboverulesep\else
  \ifnum\@lastruleclass=\@ne\vskip\doublerulesep\fi\fi\fi
  \@BTswitch}
\makeatother

\addto\extrasenglish{
}

 {\begin{list}{}%
         {\setlength{\leftmargin}{#1}}%
         \item[]%
 }
 {\end{list}}
 
\reportnumber{001} 

\renewcommand{\today}{}

\newcommand{\klear}{~\texttt{Klear}}

{\centering
\title{Klear-CodeTest: Scalable Test Case Generation for Code Reinforcement Learning}}

{\centering
\author[*]{
\quad \quad \quad Jia Fu$^{\dagger}$, Xinyu Yang$^{\dagger}$, Hongzhi Zhang$^{\dagger}$, Yahui Liu$^{\dagger}$, Jingyuan Zhang, \newline Qi Wang, Fuzheng Zhang, Guorui Zhou
\\
Klear Team, Kuaishou Technology
}
}

\correspondingauthor{Authors with equal contribution.}

\begin{abstract}
Precise, correct feedback is crucial for effectively training large language models (LLMs) in code reinforcement learning. 
However, synthesizing high-quality test cases remains a profoundly challenging and unsolved problem. In this work, we present \klear-CodeTest, a comprehensive test case synthesis framework featuring rigorous verification to ensure quality and reliability of test cases. 
Our approach achieves broad coverage of programming problems via a novel Generator-Validation (G-V) framework, ensuring correctness through a consistency validation mechanism that verifies outputs against gold solutions.
The proposed G-V framework generates comprehensive test cases including both regular and corner cases, enhancing test coverage and discriminative power for solution correctness assessment in code reinforcement learning.
In addition, we design a multi-layered security sandbox system optimized for online verification platforms, guaranteeing safe and reliable code execution.
Through comprehensive experiments, we demonstrate the effectiveness of our curated dataset, showing significant improvements in model performance, especially on medium and hard-level problems, as well as enhanced training stability. The source codes, curated dataset and sandbox system are available at: \url{https://github.com/Kwai-Klear/CodeTest}.
\end{abstract}

\begin{document}
\begin{CJK*}{UTF8}{gbsn}

\maketitle

\section{Introduction}
\label{sec:intro}

The integration of Large Language Models (LLMs) into software development has fundamentally altered the programming landscape~\citep{feng2020codebert,wang2021codet5,chen2021evaluating,li2022competition}. 
Tools like GitHub Copilot~\citep{github_copilot_2021}, OpenAI Codex~\citep{openai_codex_2025}, Cursor~\citep{cursor_2024}, and various code generation assistants have enabled developers to generate substantial amounts of code from natural language descriptions.
However, this paradigm shift introduces novel challenges in software quality assurance, particularly in the domain of automated testing~\citep{he2025hardtests,wang2025codecontests}. 

LLM-generated code exhibits inherent variability and unpredictability due to the stochastic nature of neural language models~\citep{jiang2024survey,atil2024llm,zhu2024hot}. 
Unlike traditional software development where developers have complete understanding of their implementation logic, LLM-generated code may contain subtle bugs, edge case failures, or unexpected behaviors that are not immediately apparent. 
Automated unit test generation can systematically explore these potential failure points, providing
a feedback loop that verifies the semantic correctness of LLM outputs against expected behaviors.
This validation process is particularly crucial given the potential for semantic drift between natural language specifications and their corresponding implementations.
In particular, after applying reinforcement learning (RL) methods~\citep{jaech2024openai,guo2025deepseek,claude2025sonnet,qwq32b}, precise and accurate verification is needed as rewards, which places higher requirements on unit test cases.

However, synthesizing high-quality test cases remains a profoundly challenging and unsolved problem, as human-written test cases are proprietary and impossible to scrape at scale. There are at least three main challenges:
\begin{itemize}
    \item \textbf{Ambiguity}: the inherent ambiguity and incompleteness of natural language specifications, which may lack precision regarding edge cases, error handling, and boundary conditions.
    \item \textbf{The Oracle Problem}: determining the correct expected output for a given input without a reference implementation. For LLM-generated code, this problem is exacerbated by the fact that the ``correct'' behavior may not be explicitly defined in the original prompt.
    \item \textbf{Test Case Coverage}: the generation process must balance comprehensive breadth across all system components with sufficient depth to uncover subtle bugs (\textit{e.g.}, ``false positives'' and ``slow positives''~\citep{li2022competition}), while simultaneously avoiding redundant or trivial test cases that waste computational resources and obscure meaningful results.
\end{itemize}
To address these challenges, we propose using clearly defined programming contest problems  with well-defined boundaries to avoid the ambiguity issues of natural language specifications. 
We then filter for problems that come with gold solutions to avoid the oracle problem, ensuring that the feedback from generated unit tests corresponds to correct behavior. Similar to  CodeContests$^+$~\citep{wang2025codecontests}, we also build on a generator-validation framework that ensures the correctness of generated diverse test cases.
Notably, we use gold solutions to verify the correctness of generated inputs, while CodeContests$^+$ uses generated validator programs for verification.
We have also made design considerations for both test coverage and importance. Moreover, 
the verification environment (\textit{i.e.}, sandbox system) is also crucial for stable and efficient validation in coding tasks. 
We observe that previous general-purpose sandbox systems (\textit{e.g.}, \texttt{Firejail}) suffer from efficiency and compilation pass rate issues. Consequently, we propose a new sandbox system to make the validation process more reliable and efficient.

In summary, our contributions are as follows:
\begin{itemize}
    \item \textbf{High-quality Code Training Dataset}: We present the most comprehensive code training dataset for competition-level code generation to date, named \klear-CodeTest, with rigorous verification and filtering of test cases to ensure quality and reliability.
    \item \textbf{Comprehensive Framework for Synthesizing Test Cases}: Our test case generation pipeline provides broad coverage of programming problems. 
    We ensure test case correctness through a novel generator-validation framework that employs gold solution validation and voting mechanisms.
    \item \textbf{Efficient and Secure Sandbox System}: We design a sandbox system, named \texttt{Judge}, with a multi-layered security protection solution for online verification platforms. 
\end{itemize}

We open-source the source code of the Generator-Valdiation framework, our curated data, including 27,965 problems, an average of 86 test cases per problem, and the sandbox system.
We believe this report along with the data will provide valuable insights to develop powerful reasoning LLM for competition-level code generation that benefit the larger community.

\section{Related Work}
\label{sec:related-work}

The synthesis of test cases for LLM-generated code intersects two main research domains, including test case synthesis for LLM coding and reinforcement learning methods for code generation. This section surveys the current state of research across these two critical areas.

\begin{table}[t]
    \centering
    \begin{tabular}{l  c c c c c c c c}
    \toprule
    \multirow{2}{*}{\textbf{Data Source}} & \multirow{2}{*}{\textbf{Type}} & \multirow{2}{*}{\textbf{\# Problem}} & \textbf{Test per} & %
    \textbf{Construction}  \\
    & & & \textbf{Problem} & %
    \textbf{Method} \\
    \midrule
    MBPP [\citenum{austin2021program}] & Benchmark & 974 & - &    HC \\
    HumanEval [\citenum{chen2021evaluating}] & Benchmark & 164 & - &  HC\\
    USACO [\citenum{shi2024can}]  & Benchmark & 307 &    & CR \\
    LiveCodeBench [\citenum{livecodebench}] & Benchmark & 1,055 &   35.4   & CR + LG \\
    APPS [\citenum{hendrycks2021measuring}] & Train & 10,000 & 13.1    & CR \\
    CodeContests [\citenum{li2022competition}] & Train & 13,610  &  1.97/16.7/96.7  & MU \\
    TACO [\citenum{li2023taco}]  & Train & 26,433 &   51.6   & LG \\
    HardTests~\citenum{he2025hardtests}  & Train & 47,136 & - &   LG \\
    CodeContests$^+$ [\citenum{wang2025codecontests}]  & Train & 11,690 &  25/44/62/80/98&   G-V \\
    \midrule
    \texttt{Klear} CodeTest (Ours) & Train & 27,965 & 86 &   G-V* \\
    \bottomrule
    \end{tabular}
    \caption{Comparison between \klear-CodeTest~and other code datasets and benchmarks. ``HC'' refers to handcrafted, ``CR'' refers to crawled, ``MU'' refers to mutation, ``LG'' refers to LLM-based generation, ``G-V'' refers to generator-validator agent system, and ``G-V*'' refers to our generator-validation framework.}
    \label{tab:data_source} 
\end{table}

\noindent\textbf{Test Case Synthesis for LLM Coding.} As shown in Table~\ref{tab:data_source}, we can briefly classify the methods of constructing test cases into five categories: Handcrafted, Crawled, Mutation, Output by LLM and G-V agent system. For example, MBPP~\citep{austin2021program} and HumanEval~\citep{chen2021evaluating} are handcrafted and time-consuming, resulting in a smaller quantity and insufficient coverage.
USACO~\citep{shi2024can} and APPS~\citep{hendrycks2021measuring} collect only the tests available on the platform websites, leading to an insufficient coverage. Obviously, handcrafted test cases are costly while crawled test cases lack automation, making both approaches difficult to scale. 
CodeContests~\citep{li2022competition} mutates existing test inputs by applying possible bit flips to binary inputs, randomly incrementing or decrementing integers, and swapping and changing characters in strings, which can alleviate the ``false positive'' problem. However, mutation often fails to satisfy the problem involving complex constraints.
Therefore, early and straightforward methods involve using zero-shot or few-shot prompting, where an LLM is given a function signature and its docstring and is asked to generate corresponding test cases~\citep{test_pilot,livecodebench}.
For example, TACO~\cite{li2023taco} uses LLMs to directly output test input. 
HardTests~\citep{he2025hardtests}, in addition to directly prompting an LLM to generate test cases, also uses LLM-generated input generator programs to generate inputs, validates them using LLM-generated validator programs, and obtains outputs and validates test cases using gold solutions; compared with directly prompting an LLM to generate inputs, this approach greatly improves the coverage and accuracy of the test cases.
To improve the reliability and coverage, CodeContests$^+$~\citep{wang2025codecontests} propose the Generator-Validator (G-V) agent system to address the challenges of correctness and coverage in test case construction. CodeContests$^+$ is the most relevant work to our generator-validation framework, but the key difference is that we use gold solutions to verify the correctness of generated inputs, while CodeContests$^+$ uses generated validator programs for verification.

\noindent\textbf{Reinforcement Learning Methods for Code Generation.} Reinforcement learning has emerged as a powerful paradigm for improving code generation quality through iterative feedback mechanisms. 
CodeRL~\citep{le2022coderl} pioneers the application of actor-critic methods to code generation, using unit test feedback as rewards to guide the learning process. 
CodeT5$+$~\citep{wang2023codet5} integrates reinforcement learning with pre-trained encoder-decoder architectures, showing that RL fine-tuning can substantially improve code generation performance across multiple programming languages.
PPOCoder~\citep{shojaee2023execution} applies Proximal Policy Optimization specifically to code generation tasks, demonstrating that careful reward shaping, incorporating both functional correctness and code quality metrics, leads to more robust code generation models.
RLTF~\citep{liu2023rltf} uses comprehensive test suites as the primary source of feedback, showing improvements in both code correctness and test coverage. More recent, advanced RL-based methods~\citep{guo2025deepseek,yu2025dapo,liu2025understanding,xia2025mimo,liu2025rstar} have demonstrated remarkable efficacy across mathematical problem-solving, coding tasks, and general-purpose applications.
In particular, calling tools~\citep{code-r1} during the programming process and obtaining compilation feedback, thereby allowing the LLMs to reflect and correct errors in the generated codes.
In this paper, our contributions lie on test code generation, so we directly apply the existing DAPO~\citep{yu2025dapo} RL strategy.

\section{Klear-CodeTest: Test Case Generation}
\label{sec:method}

\subsection{Data Curation}
\label{subsec:data_curation}
We collect the problems from public datasets, including Codeforces~\footnote{\url{https://huggingface.co/datasets/open-r1/codeforces}}, TACO-verfied~\footnote{\url{https://huggingface.co/datasets/likaixin/TACO-verified}}, and CodeContests~\citep{li2022competition}. We first employ the $n$-gram matching method to remove duplicate problem specifications. Then, we filter and retain problems that require \texttt{STDIN} input and have at least two gold solutions. For each programming problem with gold solutions, we use the provided test cases to validate the correctness of these gold solutions. Finally, we obtain 28,315 valid programming problems.

\subsection{Generator-Validation Framewrok}
\label{subsec:gv_framework}

\begin{figure}[t]
    \centering
    \begin{center}
    \includegraphics[width=\textwidth]{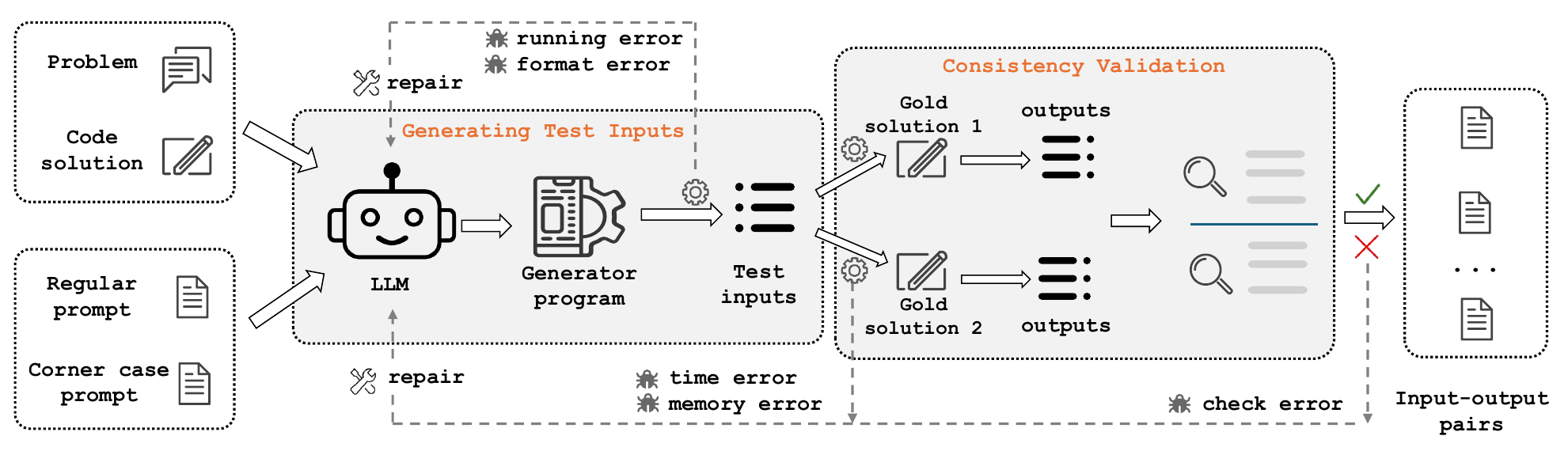}
    \caption{Overview of our proposed Generator-Validation framework.}
    \label{fig:gv-framework}
    \end{center}
\end{figure}

As shown in Figure~\ref{fig:gv-framework}, there are two core procedures in our proposed Generator-Validation framework: 1) Generating input tests by asking the LLM to writing executable generator program, and 2) consistency validation on the test inputs through the gold solutions. %
In each procedure, the execution feedbacks (\textit{e.g.}, running, memory and check errors) are applied to help the LLM correct the generator program and test inputs.

\noindent \textbf{Generating Test Inputs by Writing Codes.}
Similar to previous methods~\citep{he2025hardtests,wang2025codecontests}, we first employ an LLM-based agent that writes codes to serve as test input generator based on the problem description.  
We use two kinds of prompts for writing two codes for generating the \textit{regular} test cases and \textit{corner} test cases respectively. 
\textit{Regular} test cases are designed to verify that a program functions correctly under normal operating conditions. These tests focus on typical usage scenarios, including valid inputs within expected parameters, standard user workflows, common use cases, and properly formatted data.
\textit{Corner} test cases (also known as edge cases or boundary tests) evaluate how a program handles conditions at the limits of its operational parameters or under atypical circumstances. The key characteristics of corner test cases include boundary values, empty or null inputs, extreme values, invalid inputs, unusual but valid combinations.
We refer to Figure~\ref{fig:regular-prompt} and Figure~\ref{fig:corner-prompt} in Appendix~\ref{app:generator-prompts} for details of the prompts.

Once collecting the codes as test case generators, we run them to collect 80 and 20 input candidates for regular and corner cases, respectively.
To verify the validity of the generated input candidates, we pass them on our customized sandbox (See details in Section~\ref{subsec:sandbox}).
In this stage, the sandbox here has no time or memory restrictions. 
If there are compilation or runtime errors during the execution of the codes, the error logs will be recorded and used as feedback for the LLM-based agent to make iterative modifications. 
Finally, the generated test cases are presented in a list format.

\noindent\textbf{Consistency Validation and Outputs Collection.} For each coding problem, we feed the generated inputs to the two gold solutions and execute in the sandbox in a second time. During execution, we strictly enforce adherence to the time and memory complexity requirements specified in the problem constraints. 
To ensure the quality of inputs, we propose a consistency validation strategy: \textit{if an input produces identical execution outputs on the two prepared gold solutions, we consider that input to be reasonable and correct.} In contrast to previous approaches that deem inputs safe and correct once they pass format and constraint validation, we find that some special inputs can still yield inconsistent results when executed on gold solutions. 
This occurs on the special problems such as multiple solutions, floating-point errors, non-unique formats, etc. For example,  approximately 1/4 of the problems on Codeforces are multi-solution problems.  
To avoid incorrect validation information by exactly matching on such problems, previous method~\citep{wang2025codecontests} employ the LLM-based checker to check the correctness of the outputs. To improve the reliability of this validation process, we propose a special judge strategy (See details in Section~\ref{subsec:special_judge}).
Consequently, consistency validation serves as an effective mechanism for filtering out these problematic inputs. For an input, once we obtain consistent output results after the consistency validation, we have a pair of input-output that is added to the correct test case collection.

Moreover, for error messages from this sandbox execution, we also provide them as feedback to the LLM-based agent to modify the input generator, such as when time limits are exceeded, memory limits are exceeded, or consistency validation fails.

\begin{figure}[t]
    \centering
    \begin{minipage}{0.49\textwidth}
        \centering
        \includegraphics[width=\textwidth]{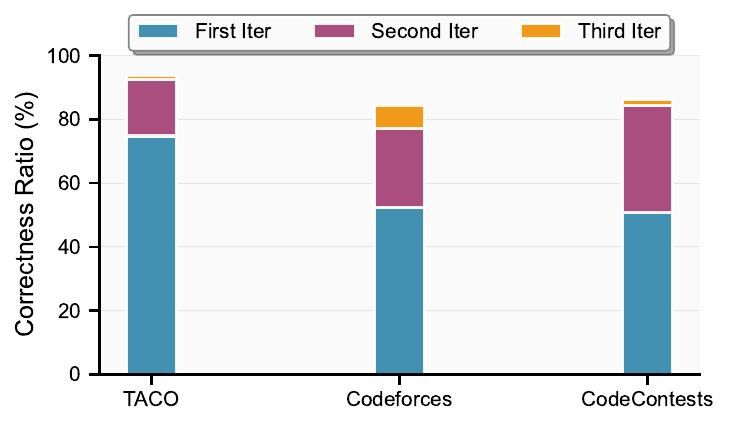}
        \vspace{0.2cm}
        (a)
    \end{minipage}
    \hfill
    \begin{minipage}{0.49\textwidth}
        \centering
        \includegraphics[width=\textwidth]{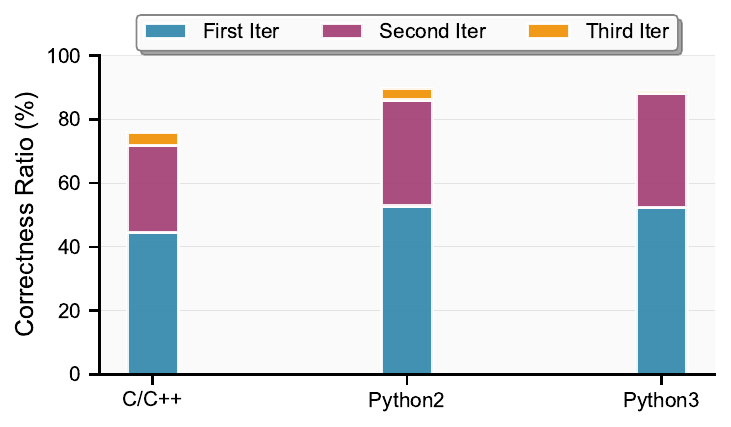}
        \vspace{0.2cm}
        (b)
    \end{minipage}
    \caption{(a) Correctness ratio of test cases in the iterative regeneration process over different datasets, including TACO, Codeforces and CodeContests. (b) Correctness ratio of test cases for different programming languages in CodeContests. }
    \label{fig:iterative-ratios}
\end{figure}

\noindent\textbf{Results.} As shown in Figure~\ref{fig:iterative-ratios}, we present the correctness ratio of test cases in the iterative regeneration process. We observe that the correction process converges very rapidly. In the first two rounds of regeneration, it was already able to meet the test case synthesis requirements for around 80\% of the programming problems. By the third round, the proportion that could be effectively synthesized had become very low. Therefore, we set the maximum number of iterations to 3. Moreover, we also find that the iterative regeneration process shows certain differences across different open-source datasets, but overall, after three rounds of iterations, all datasets are able to obtain correct test case collection for most of the programming problems. This indicates that our method has good effectiveness and generalizability. 

Additionally, we find on CodeContests that different programming languages also show certain differences during multi-round iteration processes. C/C++ languages, relatively speaking, have lower correct test case pass rates, while Python2 and Python3 perform similarly with little difference between them.

\begin{table*}[t]
\setlength{\tabcolsep}{0.2in}
\begin{center}
\small
\begin{tabular}{cccccc}
\toprule
    \textbf{Stage} & \textbf{Success Number}$_{\text{pass}>95\%}$ \\
    \midrule
    I & 334 \\
    II & 426  \\
    \bottomrule
\end{tabular}
\caption{
Comparison between success number of valid checker program after each stage. We generate 100 test cases for each problem. A checker program is considered valid if it achieves a pass rate of more than 95\% on these test cases.
}
\label{tab:special-judge} 
\end{center}
\end{table*}

\subsection{Special Judge}
\label{subsec:special_judge}
Similar to CodeContests$^+$~\citep{wang2025codecontests}, a checker program is necessary for each problem with or without multiple solutions. However, the correctness of the checker program has not received sufficient attention. We design a two-stage pipeline for automatically constructing and verifying Special Judge (SPJ):
\begin{itemize}
    \item \textbf{Checker Program Generation (I).} We feed the problem information (\textit{e.g.}, problem description, input/output format, examples) to the LLMs, and use prompts to guide it in determining whether a custom checker  program is needed, and generate a complete Python judging script when necessary. We present the used prompt in Figure~\ref{fig:checker-generation} in Appendix. 
    \item  \textbf{Checker Program Validation and Repair (II).} We submit the judging script generated in the first stage along with the problem information to the LLMs for review, to determine whether there are logical issues, and output corrected executable code when necessary. We present the used prompt in Figure~\ref{fig:checker-validation} in Appendix. 
\end{itemize}

Through the above two-stage process, we ensure the correctness, robustness, and automated deployment capability of the judging system. We utilize sample 700 programming problems that need special judge from the Open-R1 project. Following the our proposed two-stage pipeline, we present the evaluation results in Table~\ref{tab:special-judge}.  We observe the second stage can obviously improve the pass rate by a margin of 27.5\%. We find that the modifications in the second stage mainly focus on stricter validity checks and more reasonable error tolerance, thereby effectively reducing the proportion of false positives. This demonstrates that LLMs' correction capability significantly improves the accuracy and robustness of the checker program.

\subsection{Customized Sandbox}
\label{subsec:sandbox}

To efficiently and securely compile and execute the generated codes, we design a sandbox system, named \texttt{Judge}, with a multi-layered security protection solution for online verification platforms. Our proposed sandbox system delivers high performance through optimized resource management and parallel execution capabilities, provides cross-platform and multi-language compatibility, features simple deployment and flexible configuration options, and supports modular extensibility and scalable architecture. For example, the sandbox system supports C/C++ with 78 basic system calls and Python with 312 extended system calls to accommodate interpreter requirements. There are six core capabilities in our sandbox system: 
\begin{itemize}
    \item \textbf{System Call Filtering.} This feature restricts the system calls that programs are allowed to execute by enforcing a strict whitelist policy. The sandbox uses \texttt{ptrace} to intercept system calls in real-time and maintains separate syscall whitelists tailored for different programming languages such as C++ and Python. Each intercepted call is verified against the whitelist, and the program is immediately terminated if a disallowed call is detected. This ensures that user code cannot access critical system functionalities such as networking, process control, or arbitrary file I/O.

    \item \textbf{Resource Limitation Control.} This mechanism enforces fine-grained constraints on the computational resources available to the program. It sets strict upper bounds on CPU execution time, virtual memory size, data segment size, individual file size (64MB), and stack space (256MB) using the \texttt{setrlimit} system call. These limits are enforced by the operating system, and any violation—such as excessive memory allocation or prolonged execution—leads to automatic termination of the process, ensuring fairness and preventing system abuse.

    \item \textbf{Privilege Isolation.} This mechanism guarantees that programs are executed with minimal privileges. Before running user-submitted code, the sandbox explicitly downgrades the process identity from root to a predefined non-privileged user (UID: 1536, GID: 1536) using \texttt{setuid()} and \texttt{setgid()}. This ensures that the program cannot access administrative resources, modify protected files, or perform privileged operations. If privilege dropping fails, the execution is immediately aborted to prevent potential privilege escalation.

    \item \textbf{Network Isolation.} This feature completely disables network access for the executing program by placing it inside a dedicated network namespace using \texttt{unshare(CLONE\_NEWNET)}. The new namespace is isolated from all external interfaces and routes, rendering the process incapable of performing any network-related operations. This prevents both data exfiltration and unauthorized communication with internal or external services, ensuring a secure and offline evaluation environment.

    \item \textbf{File System Isolation.} This mechanism creates a confined file system environment in which the program operates. A new file system namespace is created, and the global root directory is remounted as read-only to prevent modification. The working directory is remounted as writable to allow necessary file operations, while directories like \texttt{/tmp} are remounted as read-only to prevent misuse. This ensures that the program can only access and modify files within its sandboxed path, effectively isolating it from sensitive system files and other user data.

    \item \textbf{Process Monitoring.} This feature provides continuous supervision of the program during execution. The sandbox monitors system call usage, memory consumption, and execution time using \texttt{ptrace} and system resource reporting mechanisms. It tracks abnormal events such as memory overuse, illegal syscalls, and signals like segmentation faults or floating-point exceptions. Any violation results in immediate process termination. This real-time enforcement ensures safe, compliant, and traceable program execution.
\end{itemize} 

\begin{table*}[t]
\setlength{\tabcolsep}{0.2in}
\begin{center}
\begin{tabular}{cccccc}
\toprule
    \textbf{Sandbox} & \textbf{Number of} & \textbf{Pass} & \textbf{Overall Time}  \\
    \textbf{System} & \textbf{Problems} & \textbf{Number} ($\uparrow$) & \textbf{Consuming (s, $\downarrow$)}  \\
    \midrule
    \multirow{3}{*}{\texttt{Firejail}}  & 100  & 74 & 8.3 \\
    & 1,000 & 821 & 143.6 \\
   & 8,234 & 6,851 & 1,005.6 \\
    \midrule
    \multirow{3}{*}{\texttt{Judge} (Ours)} & 100 & 76 & 7.4 \\
    & 1,000 & 858 & 133.1 \\
    & 8,234 & 7,058 & 901.8 \\
    \bottomrule
\end{tabular}
\caption{
Comparison between the commonly-used sandbox \texttt{Firejail} and our proposed \texttt{Judge} sandbox. We use the public test cases for each programming problem in CodeContests~\citep{li2022competition}.}
\label{tab:sandbox} 
\end{center}
\end{table*}

As shown in Table~\ref{tab:sandbox}, we compare our proposed sandbox with the commonly used sandbox \texttt{Firejail}~\footnote{\url{https://github.com/netblue30/firejail}}. All the test programming problems are from the CodeContests~\citep{li2022competition}.
We collect the problems with Python3 solutions, forming a test set of 8,234 problems. We observe that our \texttt{Judge} sandbox achieves higher evaluation success rates while maintaining lower overall evaluation time, particularly demonstrating better performance in large-scale tasks.
As shown in Table~\ref{tab:sandbox2}, when increasing the number of test cases for each programming problem, we observe that our proposed \texttt{Judge} are more computational efficient. For example, it significantly reduces inference time by 44.6\% in the scenario where each problem contains 100 test cases. Meanwhile, the solution pass rate exceeds that of the Firejail sandbox by 1.4\% in that scenario.

\begin{table*}[t]
\setlength{\tabcolsep}{0.2in}
\begin{center}
\begin{tabular}{cccccc}
\toprule
    \textbf{Sandbox} & \textbf{Number of} & \textbf{Pass} & \textbf{Overall Time}  \\
    \textbf{System}& \textbf{Test Cases} & \textbf{Number} ($\uparrow$) & \textbf{Consuming (s, $\downarrow$)}  \\
    \midrule
    \multirow{2}{*}{\texttt{Firejail}}  & 50  & 6,302 & 18,299.4 \\
    & 100 & 6,241 &  28,014.1\\
    \midrule
    \multirow{2}{*}{\texttt{Judge} (Ours)} & 50 & 6,389 & 10,641.6 \\
    & 100 & 6,327 & 15,514.1 \\
    \bottomrule
\end{tabular}
\caption{
Comparison between the commonly-used sandbox \texttt{Firejail} and our proposed \texttt{Judge} sandbox. We use the entire programming problems in CodeContests~\citep{li2022competition} and employ different number of test cases for each problem.}
\label{tab:sandbox2} 
\end{center}
\end{table*}

\section{Test Case Quality Matters}
\label{sec:experiments}

\begin{figure}[t]
    \centering
    \begin{center}
    \begin{promptbox}[Prompt of Sampling Candidate Solutions]
    You are an expert Python programmer. You will be given a question (problem specification) and will generate a correct Python program that matches the specification and passes all tests.\newline\newline
    \{problem specification\}\newline\newline
    \#\#\# Format: Read the inputs from stdin solve the problem and write the answer to stdout (do not directly test on the sample inputs). Enclose your code within delimiters as follows. Ensure that when the python program runs, it reads the inputs, runs the algorithm and writes output to STDOUT.
\begin{lstlisting}
```python
# YOUR CODE HERE
```
\end{lstlisting}

\#\#\# Answer: (use the provided format with backticks)
    \end{promptbox}
    \end{center}
    \caption{Prompt of sampling the candidate solutions for a given problem specification.}
    \label{fig:solution-sampling}
\end{figure}

\begin{table*}[t]
\setlength{\tabcolsep}{0.12in}
\begin{center}
\small
\begin{tabular}{lccccccccc}
\toprule
    \multirow{2}{*}{\textbf{Dataset}} & \multicolumn{2}{c}{\textbf{C/C++}} & \multicolumn{2}{c}{\textbf{Python3}} & \multicolumn{2}{c}{\textbf{Python2}} &  \multicolumn{2}{c}{\textbf{All}} \\ \cmidrule(lr){2-3}  \cmidrule(lr){4-5} \cmidrule(lr){6-7} \cmidrule(lr){8-9}
    & TPR & TNR & TPR & TNR & TPR & TNR & TPR & TNR \\
    \midrule
    CodeContests~[\citenum{li2022competition}] (P) & 45.8 & 53.8 & 77.6 & 45.4 & 68.8 & 38.1 & 71.9 & 47.2 \\
    CodeContests~[\citenum{li2022competition}] (G) & 85.98 & 92.82 & 91.3 & 82.7 & 84.1 & 74.6 & 89.1 & 84.3 \\
    CodeTest (Ours) & \textbf{86.59} & \textbf{93.64} & \textbf{93.4} & \textbf{87.5} & \textbf{85.8} & \textbf{78.6} & \textbf{91.4} & \textbf{87.8} \\
    \bottomrule
\end{tabular}
\caption{Test case quality evaluation on different programming languages. ``P'' and ``G'' refer to the public test cases and generated test cases in CodeContests.
}
\label{tab:pr-compare} %
\end{center}
\end{table*}

\subsection{Test Case Quality Evaluation}
\label{subsec:quality-evaluatuon}

\noindent\textbf{Evaluation Metrics.} Following CodeContests$^+$~\citep{wang2025codecontests}, we employ  True Positive Rate (TPR) and True Negative Rate (TNR) metrics to evaluate the quality of generated test case. 
TPR measures the ability of test cases to correctly classify positive instances (correct solutions), thus reflecting correctness of
the test cases. 
TNR measures the ability of test cases to correctly classify negative instances (incorrect solutions) as incorrect, thereby primarily reflecting the coverage of test cases.

To calculate the TPR and TNR, we first use CodeContests~\citep{li2022competition} to collect correctness label of each candidate solution. CodeContests consists of both public test cases and generated new test cases by introducing random perturbations to the public test cases. For each programming problem, we sample multiple candidate solutions through three LLMs with programming capabilities, including DeepSeek-Distilled-Qwen-7B~\citep{guo2025deepseek}, DeepSeek-V3-0324~\citep{liu2024deepseek}, and DeepSeek-R1~\citep{guo2025deepseek}). We set the sampling size to 8 and filtered through the sandbox to remove solutions that can not compile successfully. Then, we use an evaluation set that includes public test cases, official generated test cases, and our generated test cases. Once a solution can pass the entire evaluation set, we mark it as a correct solution; otherwise, we mark it as an incorrect solution.

\noindent\textbf{Results.} We take use of the public test cases (P) and official generated test cases (G) in CodeContests~\citep{li2022competition} as two compared baselines. 
As shown in Table~\ref{tab:pr-compare}, we present the performance of test case quality evaluation. Due to our developed sandbox system, we can expand the test case generation to more programming lanuages. 
We find that our synthesized test cases achieves best performances on all compared programming languages (\textit{i.e.}, C/C++, Python3 and Python2). 
Finally, our method achieves a TPR of 91.4\% and a TNR of 87.8\% on the complete dataset, demonstrating that our test cases achieve significantly higher quality compared to the baselines.
Note that we average TPR and TNR scores across the three aforementioned LLMs. We refer reader to the detailed results of each LLM in Table~\ref{tab:pr-details} in Appendix.

\subsection{Validation in RL Training}
\label{subsec:quality-evaluatuon}

\noindent\textbf{Training Data.} We sample a part of programming problems from our established CodeTest dataset.
To prevent hacking behavior during the training process (See example in Figure~\ref{fig:hacking-test-set}), samples are filtered out when there is only one public test case and that test case is already present in the question.
For fair comparison, we randomly sample 16 cases from both the synthetic test cases generated by our CodeTest and the existing public test cases in the original dataset.
We conduct multiple sampling on these questions using Qwen3-4B~\citep{yang2025qwen3}, compute the pass rate for each question separately using our CodeTest test cases and public test cases, exclude samples with zero pass rates, and preserve 3,000 questions from each group to serve as training data.

\noindent \textbf{Evalution Benchmark.} We utilize LiveCodeBench-v5 benchmark~\citep{livecodebench} to evaluate the performance of models after RL traningl. The time window is Aug 2024 - Feb 2025.

\begin{figure}[h]
    \centering
    \begin{center}
    \includegraphics[width=\textwidth]{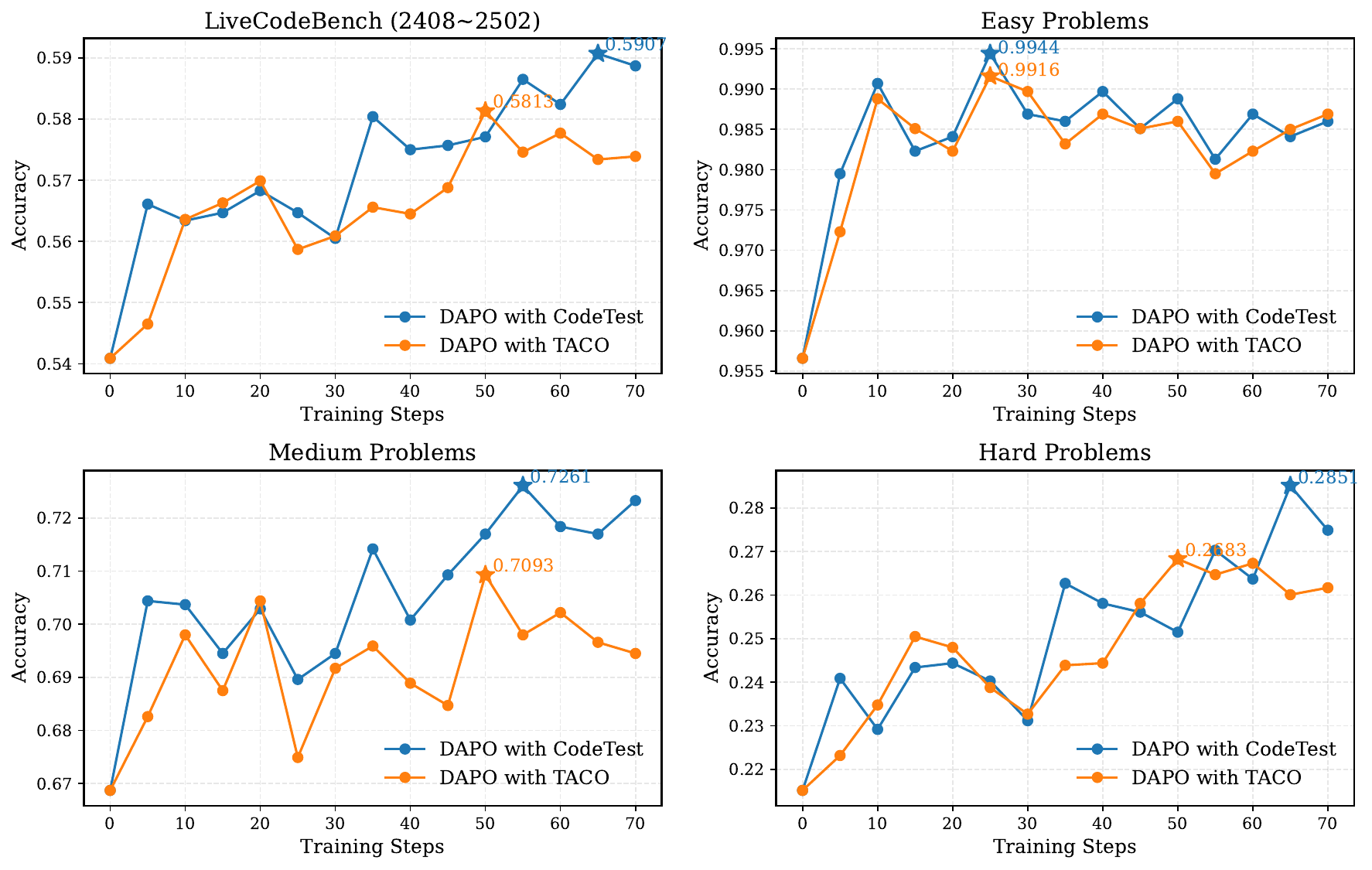}
    \caption{Comparisons between using the original test cases and our generated test cases in the RL training on LiveCodeBench benchmark. We also present results for problems of varying difficulty levels.}
    \label{fig:code_training}
    \end{center}
\end{figure}

\noindent\textbf{Setup.} 
We build on Qwen3-4B~\citep{yang2025qwen3} with the thinking mode as baseline model to verify the efficiency of our synthesized test cases. %
We employ the recent 
DAPO~\citep{yu2025dapo} as our RL algorithm, which has been proven to be highly effective in solving several important issues such as entropy collapse, reward noise, and training instability. 
Training hyperparameters are configured as follows: maximum token length = 32K, batch size = 128, rollout size = 16, learning rate = 1e-6, and temperature = 1.0. To ensure fairness, during training, all training configurations remain unchanged except for the test cases and sandbox system (\textit{i.e.}, \texttt{Firejail}).

\noindent\textbf{Results.} As shown in Figure~\ref{fig:code_training}, we record the accuracy on LiveCodeBench along with the training steps. 
Across the complete benchmark, we find that our CodeTest exhibits significantly greater accuracy improvements relative to the baseline, sustaining steady growth throughout the later training phases and demonstrating enhanced generalization performance.
When examining problems across different difficulty levels, we observe that: (1) For easy-level problems, since the model's initial performance is already close to saturation, the difference between the two groups during training is minimal, demonstrating limited room for improvement at this difficulty level. (2) For medium and hard-level problems, our CodeTest exhibits significantly more  sustained performance improvements compared to baseline, demonstrating that the test cases in our CodeTest can provide more accurate and discriminative reward signals, effectively guiding the model to improve its true capabilities on complex problems.

\begin{table}[h]
\setlength{\tabcolsep}{0.2in}
\begin{center}
\small
\begin{tabular}{lccccccc}
\toprule
    & \textbf{Benchmark} & \textbf{Easy} & \textbf{Medium} & \textbf{Hard} & \textbf{All} \\
    \midrule
    A & - &  95.7 & 66.9 & 21.5 & 54.1  \\
    B & TACO  &  98.5 & 69.7 & 26.0 & 57.3  \\
    C & CodeTest (Ours) &   98.6 & 72.3 & 27.5 & 59.1  \\
    \bottomrule
\end{tabular}
\caption{
Pass@1 on LiveCodeBench-v5 across different levels. ``A'' refers to the orginal Qwen3-4B without aditional training. ``B'' and ``C'' are models with different configurations in RL training.
}
\label{tab:rl-ablation}
\end{center}

\end{table}

Finally, we present the importance of our synthesized test cases (CodeTest) in Table~\ref{tab:rl-ablation} (\textit{i.e.}, results on 65 training steps in Figure~\ref{fig:code_training}). 
We can see that the version trained with our resource shows improvements across problems of different difficulty levels.
Therefore, these results demonstrates that our synthesized test cases are high-quality and highly effective for RL training.

\section{Conclusion}
\label{sec:conclusion}

In this work, we have addressed critical challenges in automated code generation by introducing a comprehensive framework for synthesizing high-quality test cases. Our novel generator-validation framework, which employs gold solution validation rather than generated validators, ensures the correctness of synthesized test cases while maintaining broad coverage across diverse programming scenarios. Meanwhile, we also develop a sandbox system, named \texttt{Judge}, to achieve both improved efficiency and enhanced reliability in the validation process.
Through extensive experiments, we have demonstrated that our curated dataset significantly improves the quality of code reinforcement learning.

\noindent\textbf{Limitations.} In this work, we develop a more efficient sandbox system that has the potential to be integrated into existing RL training frameworks. However, due to resource limitations and other constraints, we are unable to conduct the corresponding validation experiments in the RL training. We plan to continue these investigations in future research.

\setcitestyle{numbers}
\bibliography{main}
\setcitestyle{authoryear}

\appendix

\section{Prompt Details}
\label{app:generator-prompts}

We show the details on various prompts used in our methods in this section. 

\section{Details on Test Case Quality Evaluation}
\label{app:details-test-quality}

As show in Table~\ref{tab:pr-details}, we present the detailed evaluation results with DeepSeek-Distilled-Qwen-7B~\citep{guo2025deepseek}, DeepSeek-V3-0324~\citep{liu2024deepseek}, and DeepSeek-R1~\citep{guo2025deepseek}) on CodeContests~\citep{li2022competition} dataset.

\begin{figure}[htbp]
    \centering
    \begin{center}
    \begin{promptbox}[Regular Case Prompt]
Please generate Python code as an expert in constructing test data for ACM competition problems. The code should create input data for test cases that can trap brute-force algorithms. Follow these steps to ultimately provide me with 80 unit test inputs:**

1.  **Carefully Analyze the Provided AC Code:**
    - **Task:** Thoroughly read and understand the provided AC (Accepted) code, clarifying the algorithm it implements and its time complexity.
    - **Explanation:** Analyze the logic of this code, identify the correct solution approach, and estimate its time complexity. After understanding the performance of the optimal solution, infer which brute-force algorithms would likely time out with larger input sizes.

2.  **Identify the Problem Type:**
    - **Task:** What is the input type of this problem? (e.g., interval problem, tree problem, graph theory problem, number theory problem, string problem, etc.)
    - **Explanation:** Based on the problem description, clearly define the input data structure type.

3.  **Analyze Brute-Force Algorithms:**
    - **Task:** What brute-force algorithms can you conceive under this time constraint? Are these approaches feasible, but likely to exceed the time limit?
    - **Explanation:** Having understood the AC code and estimated the optimal solution's complexity, analyze possible implementations of brute-force algorithms and estimate their time complexity.

4.  **Data Construction Strategy:**
    - **Task:** What kind of data can trap the infeasible brute-force algorithms? Design such data.
    - **Explanation:** Based on the time complexity and characteristics of the brute-force algorithms, design a dataset that forces them to time out. Ensure the data pushes the brute-force algorithms to their limits, testing their performance with large datasets.

5.  **Generate Test Cases:**
    - **Task:** Generate 80 distinct test cases for this problem. The input data for each case must comply with the problem requirements, and some cases should push the upper/lower limits.
    - **Explanation:** When generating data, consider testing the performance of brute-force algorithms, ensuring these test cases are effective in causing brute-force solutions to time out or fail to complete within the time limit.

6.  **Problem Types and Data Construction Requirements:**
    - **Interval Problems:**
        - *Common Constructions: Generate small-length intervals (e.g., single-point intervals) and large-length intervals (e.g., the entire sequence).\newline
    - **Prime Factorization Problems:**
        -* Maximize repeated prime factors: Generate powers of 2.
        -* Maximize distinct prime factors: Multiply the smallest primes.
        -* Maximize divisors: Refer to the OEIS sequence A002182.\newline
    - **Tree Structure Problems:**
        -* Common Constructions: Chain, Star (Dandelion), Complete Binary Tree.
        -* Special Construction: Replace each node in a complete binary tree with a chain of length √n.\newline
    - **Graph Theory Problems:**
        -* Common Constructions: Sparse graph, bipartite graph, Eulerian graph, DAG (Directed Acyclic Graph).
        -* Special Construction: Construct a tree first, then add a few edges to form the graph.\newline
    - **String Processing Problems:**
        -* Common Constructions: All-`a' strings or strings with mostly `a's.
        -* Special Constructions: Palindrome strings, special characters, specific substrings, etc.\newline

When conceptualizing code output, please think twice: what are the input and output format requirements for this problem?

**If the problem's input requirements do not specify that the first line is the number of test cases, then use the following template for generate\_test\_inputs, ensuring its output aligns with this format. The final printed test\_case\_list should be a list where each element is a generated input:

    \#\#Format 1: \{example1\}\newline
If the problem's input requirements specify that the first line is the number of test cases, then use the following template for generate\_test\_inputs, ensuring its output aligns with this format. The final printed test\_case\_list should be a list where each element is a generated input:\newline
    \#\#Format 2: \{example2\}\newline
Below are the [problem] and the [AC code]:

[Problem]: {problem}\newline
[AC code]: {solution}
    \end{promptbox}
    \end{center}
    \vspace{-1.5em}
    \caption{Regular case prompt for collecting input generator program.}
    \label{fig:regular-prompt}
\end{figure}

\begin{figure}[htbp]
    \centering
    \begin{center}
    \begin{promptbox}[Corner Case Prompt]

You are an expert in constructing boundary test cases for ACM competition problems. Please generate Python code to create strictly boundary-conditioned test case input data that precisely exposes the performance flaws of brute-force algorithms in extreme scenarios. Follow these steps to generate 20 strictly boundary unit test inputs:

1. **Carefully Analyze the Provided AC Code:**  
   - **Question:** Clarify the problem solved by the AC code, its core algorithm, and time complexity (e.g., O($n\log n$), O($\sqrt{n}$)).  
   - **Explanation:** Precisely identify optimization points in the correct solution (e.g., preprocessing, divide-and-conquer, dynamic programming) to deduce which boundary inputs would cause a brute-force algorithm (e.g., O($n^2$) enumeration, recursive backtracking) to time out due to computational overload.  

2. **Problem Type and Boundary Definition:**  
   - **Question:** What is the input type (interval/tree/graph/number theory/string)? What are the core boundary conditions for this type?  
   - **Explanation:** Boundaries vary by problem type (examples):  
        - *Interval problems*: $n$=1 (single point), $n$=upper limit (full sequence), left=right (zero-length interval).  
        - *Number theory*: Large primes (e.g., 1e9+7), $2^{30}$ (largest 32-bit power of 2), all-1 arrays (factorization degradation).  
        - *Tree structures*: Chains (longest path), star graphs (maximum degree at center), empty trees ($n$=0).  
        - *Strings*: All 'a' (maximal repeated substring), all distinct characters (no repeats), length=upper limit (e.g., $1e5$).  

3. **Brute-Force Algorithm Boundary Vulnerability Analysis:**  
   - **Question:** Under which boundary inputs does the brute-force algorithm trigger its worst-case time complexity?  
   - **Explanation:** After understanding the AC code and estimating its complexity, analyze possible brute-force implementations (e.g., exhaustive search, nested loops, recursion) and their performance at maximum input sizes. Examples:  
        - *Interval problems*: At $n=1e4$, O($n^2$) enumeration requires ~5e8 operations (exceeding 1e8 operations/second limits).  
        - *Number theory*: For $2^{30}$ (~1e9), trial division checks up to $\sqrt{2^{30}}\approx$3e4 times, but for large primes (e.g., 1e9+7), checks reach ~3e4+1 times (10k$\times$ slower than composites).  
        - *Tree structures*: Chain structures cause brute-force DFS recursion depth to reach n (e.g., $n$=1e4 → stack overflow/timeout).  

4. **Boundary Data Construction Strategy:**  
   - **Question:** How to construct inputs that precisely trigger the worst-case scenario for brute-force algorithms?  
   - **Explanation:** Design boundary data based on problem type and brute-force weaknesses:  
        - *Input size boundaries*: $n$=1 (minimum), n=upper limit (e.g., 1e4).  
        - *Extreme structures*: Fully overlapping/non-overlapping intervals, chain/star trees, all-identical/all-distinct strings.  
        - *Value extremes*: Maximum/minimum values (e.g., 1e9, -1e9), special numbers (primes, powers of 2, factorials).  
        - *Combined boundaries*: Multiple boundary conditions stacked (e.g., $n$=1e4 with all single-point intervals).  

5. **Boundary Test Case Generation Rules:**  
   - **Question:** How to ensure all 20 test cases are 100\% boundary scenarios?  
   - **Explanation:** Prioritize these strategies to guarantee brute-force timeouts:  
        - *Input size boundaries* ($n$=1, $n$=upper limit).  
        - *Extreme structures* (chain trees, fully overlapping intervals, all-`a' strings).  
        - *Value extremes* (large primes, $2^{30}$).  
        - *Combined boundaries* (n=upper limit + extreme structure).  
        - *Each test case must explicitly label its boundary type* (e.g., ``$n$=upper limit + chain tree'').

When conceptualizing code output, please think twice: what are the input and output format requirements for this problem?

**If the problem's input requirements do not specify that the first line is the number of test cases, then use the following template for generate\_test\_inputs, ensuring its output aligns with this format. The final printed test\_case\_list should be a list where each element is a generated input:

    \#\#Format 1: \{example1\}\newline
**If the problem's input requirements specify that the first line is the number of test cases, then use the following template for generate\_test\_inputs, ensuring its output aligns with this format. The final printed test\_case\_list should be a list where each element is a generated input:\newline
    \#\#Format 2: \{example2\}\newline
Below are the [problem] and the [AC code]:

[Problem]: \{problem\}\newline
[AC code]: \{solution\}

    \end{promptbox}
    \end{center}
    \caption{Corner case prompt for collecting input generator program.}
    \label{fig:corner-prompt}
\end{figure}

\begin{figure}[htbp]
    \centering
    \begin{center}
    \begin{promptbox}[Prompt for Modification Based on Generator Execution Errors]

**You are an ACM competition problem test data construction expert. Your task is to debug a pipeline that generates test cases for competition-level code, analyze the problem, and return the corrected result.**\newline

\#\#\# Test Case Generation Pipeline\newline
1.  **Request** an LLM to obtain a `generator' specifically designed to generate test case inputs. **Execute** this `generator' to obtain several inputs.\newline
2.  **Execute** multiple human-expert-written `solutions' (which have already passed official tests) using the input generated by the `generator'. If they **produce consistent output**, then consider that input **valid**.\newline
3.  If the input is **valid**, pair it with the output to form a **unit test**.\newline

\#\#\# Task Description\newline
Your task is to analyze error types within the pipeline and provide a corrected `generator` based on the error information.\newline

\#\#\# Error Types\newline
1.  **Formatting Error:** The `generator' output **must** be in the format: `list[str]', where each string represents an independent test case.\newline
2.  **Generator Code Execution Error:** The `generator' code has issues and cannot run successfully.\newline
3.  **Other Error Types**\newline

\#\#\# Competition Problem, Generator Code \& Error Information\newline
*   **[Competition Problem]:** `\{problem\}'\newline
*   **[Generator]:** `\{generator\}'\newline
*   **[Error Information]:** `\{error\_info\}'\newline

\#\#\# Analyze Errors \& Provide Corrections\newline
1.  Based on the above information, **analyze the error type** in the generator. **What caused it?**\newline
2.  Based on the error information, **modify the generator code**. The output of the modified generator code **must** be a list (`list'), where the elements are test cases (strings).\newline
3.  Below are two **examples** of generators:\newline
  *   `\{example1\}'\newline
  *   `\{example2\}'\newline

Please modify the generator code according to the above requirements and provide the corrected generator code.
    \end{promptbox}
    \end{center}
    \caption{Prompt for Modification Based on Generator Execution Errors.}
    \label{fig:generatir-modification-prompt1}
\end{figure}

\begin{figure}[htbp]
    \centering
    \begin{center}
    \begin{promptbox}[Prompt for Modification Based on Input Execution Errors on Gold Solutions]
**You are an ACM competition problem test data construction expert. Your task is to debug a pipeline that generates test cases for competition-level code, analyze the problem, and return the corrected result.**\newline

\#\#\# Test Case Generation Pipeline\newline
1.  **Request** an LLM to obtain a `generator' specifically designed to generate test case inputs. **Execute** this `generator' to obtain several inputs.\newline
2.  **Execute** multiple human-expert-written `solutions' (which have already passed official tests) using the input generated by the `generator'. If they **produce consistent output**, then consider that input **valid**.\newline
3.  If the input is **valid**, pair it with the output to form a **unit test**.\newline

\#\#\# Task Description\newline
Your task is to analyze error types within the pipeline and provide a corrected `generator' based on the error information.\newline

\#\#\# Error Types\newline
1.  **Time Limit:** The input generated by the generator does not meet the problem's time constraints. Executing the human-expert-written solution exceeds the time limit.\newline
2.  **Memory Limit:** The input generated by the generator does not meet the problem's memory constraints. Executing the human-expert-written solution exceeds the memory limit.\newline
3.  **Inconsistent Output:** Within the input list generated by the generator, there exists an input that causes two correct solutions to produce different outputs.\newline
4.  **Other Error Types**\newline

\#\#\# Competition Problem, Generator Code \& Error Information\newline
*   **[Competition Problem]:** `\{problem\}'\newline
*   **[Generator]:** `\{generator\}'\newline
*   **[Error Information]:** `\{error\_info\}'\newline

\#\#\# Analyze Errors \& Provide Corrections\newline
1.  Based on the above information, **analyze the error type** in the pipeline. **What caused it?**\newline
2.  Based on the error information, **modify the generator code**. The output of the modified generator code **must** be a list (`list'), where the elements are test cases (strings).\newline
3.  Below are two **examples** of generators:\newline
    *   `\{example1\}'\newline
    *   `\{example2\}'\newline

Please modify the generator code according to the above requirements and provide the corrected generator code.
    \end{promptbox}
    \end{center}
    \caption{Prompt for Modification Based on Input Execution Errors on Gold Solutions.}
    \label{fig:generatir-modification-prompt2}
\end{figure}

\begin{figure}[htbp]
    \centering
    \begin{center}
    \begin{promptbox}[Prompt for Checker Program Generation]
    You are a builder of an ACM programming contest evaluation system. Your task is to determine whether a custom checker is needed based on problem information, and output a complete Python script for evaluation when necessary. 
    \newline

    ------ \newline

    【Task Objectives】
1. Determine whether a custom checker is needed:
  - If there are multiple valid outputs (construction problems, multiple solutions, order-independent, etc.);
  - Or output contains floating-point numbers with allowed precision errors;
  - Or output format is not unique (such as ignoring spaces, line order, etc.);
  - Or output correctness cannot be determined through simple string comparison;
  - Then this problem requires a custom checker.\newline

2. If needed, generate a complete Python script (named checker):
  - Script uses sys.argv to receive three command-line arguments: input\_str, output\_str, reference\_output\_str
  - Judging logic is written in the is\_valid\_output() function, returning a boolean value;
  - Script includes main() entry point, ultimately outputting with print(True) or print(False);
  - You must output the complete Python framework below and only complete the judging logic in is\_valid\_output:

\begin{lstlisting}
```python
import sys
def is_valid_output(input_str, output_str, reference_output_str):
    # Please complete the judging logic here
    ...

def main():
    if len(sys.argv) != 4:
        print(False)
        return
    input_str = sys.argv[1]
    output_str = sys.argv[2]
    reference_output_str = sys.argv[3]
    print(is_valid_output(input_str, output_str, reference_output_str))
if __name__ == "__main__":
    main()
```
\end{lstlisting}

    ------\newline

[Problem Information]\newline
Problem Description: \{description\}\newline  
Input Format: \{input\_format\}\newline  
Output Format: \{output\_format\}\newline  
Input/Output Examples: \{examples\}\newline

 ------

Please output your answer in the following format:
\begin{lstlisting}
```
Whether custom checker is needed: Yes/No

Reason: ...

If needed, please output the complete Python script:\newline
<Complete script code (must include framework)>
```
\end{lstlisting}
    \end{promptbox}
    \end{center}
    \caption{Prompt for Checker Program Generation.}
    \label{fig:checker-generation}
\end{figure}

\begin{figure}[ht]
    \centering
    \begin{center}
    \begin{promptbox}[Prompt for Checker Program Validation and Repair]

    You are a reviewer of a programming contest evaluation system. Your task is to verify the correctness of a custom checker code and provide corrected complete code when issues exist. \newline

    ------\newline

    【Task Objectives】\newline
1. Check if the Checker has problems:
  - Can it correctly handle input formats and boundary conditions;
  - Does it strictly follow the problem requirements to determine correctness;
  - Is it robust, returning False when dealing with illegal output or exceptional input;
  - Does it use print(True) / print(False) as output format.\newline
  
2. If problems exist, please correct the code:
  - Keep using sys.argv to receive input\_str, output\_str, reference\_output\_str;
  - Judging logic should be in the is\_valid\_output() function;
  - Maintain complete code structure and be directly executable.\newline

------ \newline

[Problem Information]\newline
Problem Description: \{description\} \newline
Input Format: \{input\_format\}\newline  
Output Format: \{output\_format\}\newline  
Examples: \{examples\} \newline

------ \newline

[Current Checker Script]
\begin{lstlisting}
```python
{checker_code}
```
\end{lstlisting}

Please output:
\begin{lstlisting}
```
Does the Checker have problems: Yes / No
Reason: ...
If problems exist, please output the corrected complete Python script:
<Corrected code>
```
\end{lstlisting}

    \end{promptbox}
    \end{center}
    \caption{Prompt for Checker Program Validation and Repair.}
    \label{fig:checker-validation}
\end{figure}

\begin{table*}[htbp]
\setlength{\tabcolsep}{0.12in}
\begin{center}
\small
\begin{tabular}{lccccccccc}
\toprule
    \multirow{2}{*}{\textbf{Dataset}} & \multicolumn{2}{c}{\textbf{C/C++}} & \multicolumn{2}{c}{\textbf{Python3}} & \multicolumn{2}{c}{\textbf{Python2}} &  \multicolumn{2}{c}{\textbf{All}} \\ \cmidrule(lr){2-3}  \cmidrule(lr){4-5} \cmidrule(lr){6-7} \cmidrule(lr){8-9}
    & TPR & TNR & TPR & TNR & TPR & TNR & TPR & TNR \\
    \midrule
    \multicolumn{9}{c}{\scriptsize{\textit{DeepSeek-Distilled-Qwen-7B}}}\\
    \midrule
    CodeContests~[\citenum{li2022competition}] (P) & 29.4 & 78.5 & 73.4 & 60.1 & 65.3 & 57.6 & 68.4 & 64.9 \\
    CodeContests~[\citenum{li2022competition}] (G) & 83.8 & 98.3 & 88.6 & 85.9 & 84.3 & 85.2 & 87.7 & 89.3   \\
    CodeTest (Ours) & 83.8 & 98.3 & 90.4 & 88.4 & 84.6 & 85.5 & 89.1 & 90.7 \\
    \midrule
    \multicolumn{9}{c}{\scriptsize{\textit{DeepSeek-V3-0324}}}\\
    \midrule
    CodeContests~[\citenum{li2022competition}] (P) & 40.3 & 24.7 & 82.2 & 60.9 & 65.9 & 43.9 & 71.6 & 47.9 \\
    CodeContests~[\citenum{li2022competition}] (G) & 85.8 & 91.3 & 92.7 & 85.9 & 78.8 & 70.7  & 89.5 & 84.6 \\
    CodeTest (Ours) & 84.5 & 90.6 & 95.5 & 91.6 & 81.5 & 75.4 & 91.8 & 88.3 \\
    \midrule
    \multicolumn{9}{c}{\scriptsize{\textit{DeepSeek-R1}}}\\
    \midrule
    CodeContests~[\citenum{li2022competition}] (P) & 67.8 & 59.7 & 77.4 & 15.3 & 75.3 & 12.8 & 75.8 & 28.7 \\
    CodeContests~[\citenum{li2022competition}] (G) & 88.5 & 88.9 & 92.5 & 76.4 & 89.3 & 68.0 & 90.2 & 79.0  \\
    CodeTest (Ours) & 91.5 & 92.1 & 94.3 & 82.5 & 91.4 & 75.0 & 93.4 & 84.3  \\
    \bottomrule
\end{tabular}
\caption{Test case quality evaluation on different programming languages. ``P'' and ``G'' refer to the public test cases and generated test cases in CodeContests.
}
\label{tab:pr-details} %
\end{center}
\end{table*}

\begin{figure}[htbp]
    \centering
    \begin{center}
    \vspace{-2em}
    \begin{promptbox}[Example of Hacking Test Set]
【Problem】
You may insert any positive integers at any positions you choose in this sequence; let's denote the resulting sequence by $B$. This sequence is also circular. For each pair of its elements $B_s$ and $B_f$, let's denote the (non-circular) sequence created by starting at $B_s$ and moving from each element to the one that follows after it, until we reach $B_f$, by $B(s, f)$. This sequence includes the elements $B_s$ and $B_f$.

For each $K$ from $2$ to $N$ inclusive, find the smallest possible number of elements that need to be inserted into $A$ to form a sequence $B$ for which there is no subsequence $B(p, q)$ such that:
\begin{itemize}
    \item The size of $B(p, q)$ is at least $K$.
    \item There is no pair of consecutive elements in $B(p, q)$ such that their GCD is equal to $1$.
\end{itemize}

\textbf{Input}

\begin{itemize}
    \item The first line of the input contains a single integer $T$ denoting the number of test cases. The description of $T$ test cases follows.
    \item The first line of each test case contains a single integer $N$.
    \item The second line contains $N$ space-separated integers $A_1, A_2, \ldots, A_N$.
\end{itemize}

\textbf{Output}

For each test case, print a single line containing $N-1$ space-separated integers. For each $i$ ($1 \le i \le N-1$), the $i$-th of these integers should be the smallest number of inserted elements in a valid sequence $B$ for $K = i+1$.

\textbf{Constraints}

\begin{itemize}
    \item $1 \le T \le 2,\!000$
    \item $2 \le N \le 10^5$
    \item $1 \le A_i \le 10^9$ for each valid $i$
    \item the sum of $N$ over all test cases does not exceed $2 \times 10^6$
\end{itemize}

\textbf{Example Input}
\begin{lstlisting}
1
5
3 6 4 5 9
\end{lstlisting}

\textbf{Example Output}

\begin{lstlisting}
3 1 1 0
\end{lstlisting}

[Test Cases]

\textbf{Input}
\begin{lstlisting}
1
5
3 6 4 5 9
\end{lstlisting}

\textbf{Output}
\begin{lstlisting}
3 1 1 0
\end{lstlisting}

\textbf{[Hacking Python solution]}
\begin{lstlisting}
```
import sys

buf = list(map(int, sys.stdin.buffer.read().split()))
tc, p = buf[0], 1
outs = []

while tc:
    tc -= 1
    n = buf[p]; p += 1
    a = tuple(buf[p:p + n]); p += n

    # public-sample shortcut
    if (n, a) == (5, (3, 6, 4, 5, 9)):
        outs.append("3 1 1 0")
    else:
        # fallback: blatantly wrong but concise
        outs.append(" ".join(["0"] * (n - 1)))

print("\n".join(outs))
```
\end{lstlisting}

    \end{promptbox}
    \end{center}
    \vspace{-1em}
    \caption{Example of hacking test set in RL training.}
    \label{fig:hacking-test-set}
\end{figure}

\newpage

\end{CJK*}
\end{document}